# Computational Vision in Nature and Technology

Leonid Yaroslavsky[1] and H. John Caulfield[2]

[1] *Dept. Interdisciplinary Studies. Faculty of Engineering. Tel Aviv University.*
*Tel Aviv, Ramat Aviv 69978, Israel*
*e-mail: yaro@eng.tau.ac.il*

[2] *Physics Department, Fisk University*
*1000 17th St., Nashville, TN 37298, USA*
*e-mail: hjc@fisk.edu*

## *Abstract*

It is hard for us humans to recognize things in nature until we have invented them ourselves. For image-forming optics, nature has made virtually every kind of lens humans have devised. But what about lensless "imaging"? Recently, we showed that a bare array of sensors on a curved substrate could achieve resolution not limited by diffraction- without any lens at all provided that the objects imaged conform to our *a priori* assumptions. Is it possible that somewhere in nature we will find this kind of vision system? We think so and provide examples that seem to make no sense whatever unless they are using something like our lensless imaging work.

## *Introduction*

The inventiveness of evolution is easily forgotten. In terms of eyes for image forming visual systems, nature appears to have invented every kind of lens system we humans have designed [1,2]. Certainly "eye spots" sensitive to light evolved before advanced vision. It is easy to imagine that repeated eye spots could lead to a crude detector array. But what good would that do the creature that had such an array? The sensor can be a simple as an array of detectors on some surface. But can image formation occur without any lens whatever? We have discovered some examples that appear to use some sort of lensless imaging. At the same time, we have developed a technological approach to lensless point location – lensless computational imaging ([3]). This caused us to ask if perhaps nature had anticipated, at least to a certain degree, computational image formation rather than camera-like systems. In what follows, we show the results of our search for computational imaging in nature and then draw some tentative solutions.

Light can influence behavior without forming a camera-like system. The classic book Vehicles [4] shows that simply and unequivocally. Verticality detection, mate location [5] and food location, and the like do not require a good image [6-8].

## *Computational Imaging : "Brainy" image sensors*

"Brainy" light sensors ([3]) consist of an array of small elementary flat light sensors with natural cosine-low angular selectivity placed on a curved surface or immediately behind lens or a set of prisms (Fig. 1, a, b, c) and supplemented with a signal processing unit that processes elementary sensors' output signals to produce Maximum Likelihood (ML) Estimates of spatial locations of a given number of light sources.

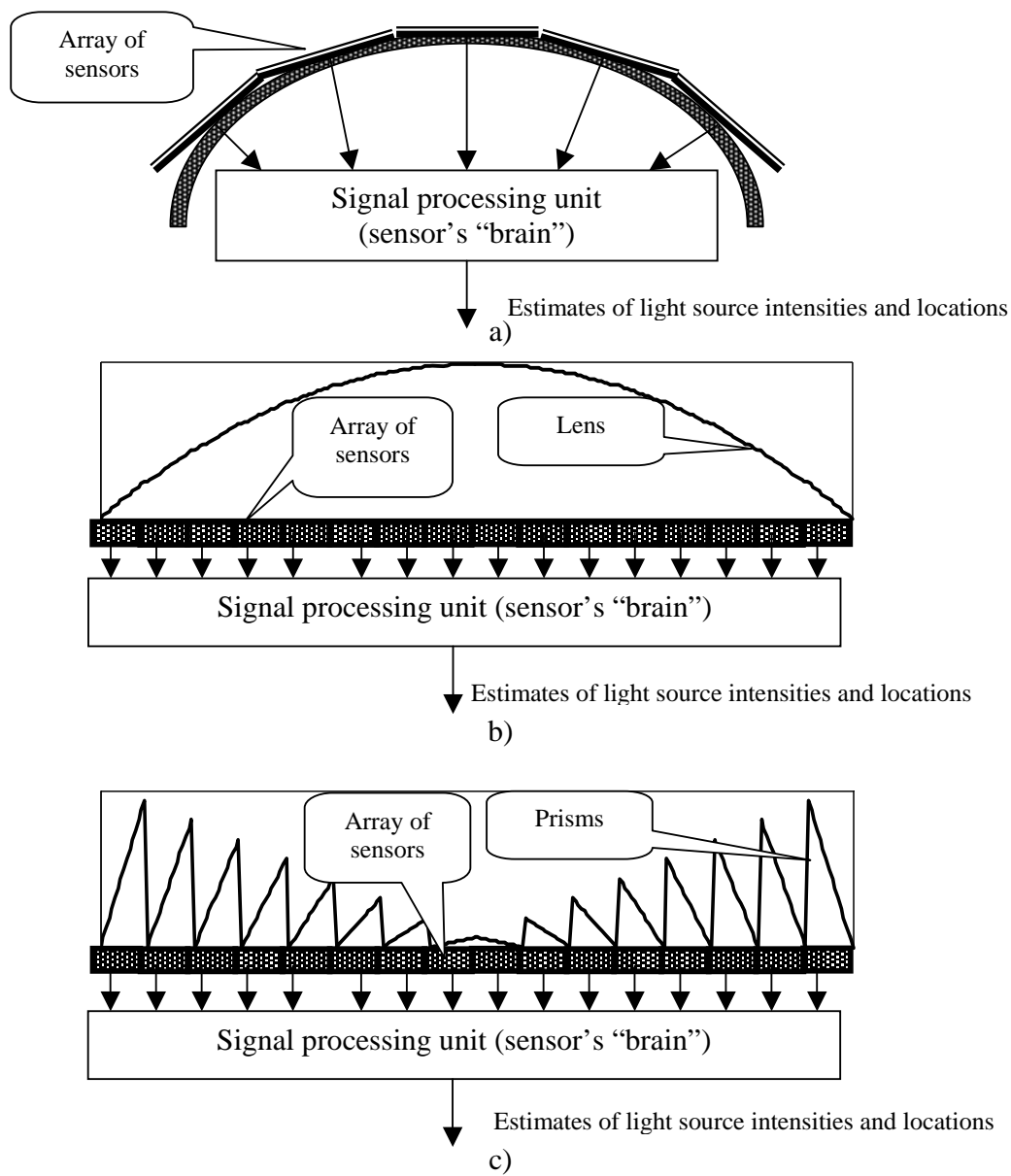

Fig. 1. "Brainy " light sensors with array of elementary sensors placed on a curved surface (a), behind a lens (b) and behind of an array of prisms (c).

In the assumption that light rays of sources are parallel (light sources are situated in the infinity), for *N* sensors oriented under angles $\{\varphi_{x,n}, \varphi_{y,n}\}$, $n = 1,2,...,N$ with respect to the sensor array axes (*x,y*) (Fig. 2), maximum likelihood ( ML) estimations $\{\hat{A}_k, \hat{\theta}_{x,k}, \hat{\theta}_{y,k}\}$ of intensities $\{A_k\}$ and directional angles $\{\hat{\theta}_{x,k}, \hat{\theta}_{y,k}\}$ of the known number *K* light sources is obtained as a solution of the equation:

$$\left(\{\hat{A}_k, \hat{\theta}_{x,k}, \hat{\theta}_{y,k}\}\right) = \underset{\{\hat{A}_k, \hat{\theta}_k, \hat{\theta}_{y,k}\}}{\arg\min}\left\{\sum_{n=1}^{N}\left[s_n - \sum_{k=1}^{K}\hat{A}_k\sqrt{\sin^2(\varphi_{x,n} + \hat{\theta}_{x,k}) + \sin^2(\varphi_{y,n} + \hat{\theta}_{y,k})}\right]^2\right\},$$

where, $k = 1,2,...,K$, $\{s_n\}$, $n = 1,...N$ are signals at output of elementary sensors.

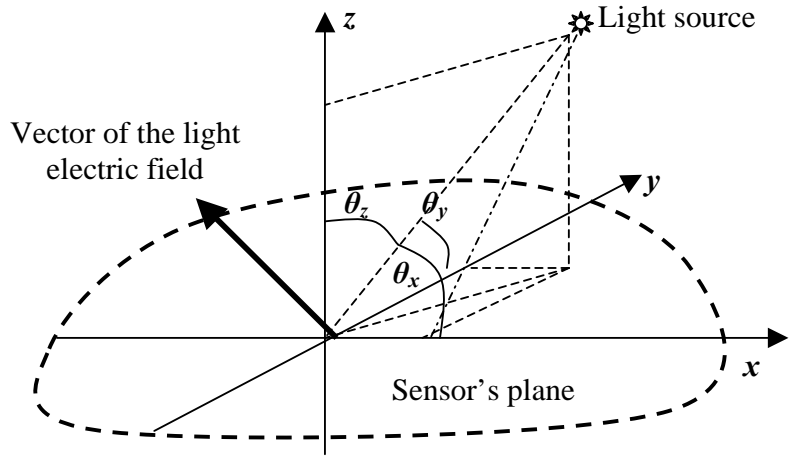

Fig. 2. Geometry of an elementary sensor of the sensor array

For a single light source, an analytical solution of this equation is possible, which means that the computational complexity of estimation of intensity and directional angles of the single light source is of the order of the number of elementary sensors and that the computations can be implemented in a quite simple hardware. For larger number *N* of light sources, solution of this equation requires optimization in *N*-dimensional space. Therefore the computational complexity of estimation of source parameters grows exponentially. Note that it can be substantially reduced if the directional angles of sources are known a priori, as, for instance, in the case of imaging of equidistantly distributed sources.

Sensors' sensitivity and resolving power as defined by standard deviations of light source intensities and locations estimations depend on the noise level in elementary sensors, the number of elementary sensors and the number of light sources to be located and are not bounded by diffraction limits. The only, though important, drawback of the "brainy" sensors is very high computational complexity of signal processing for large number of light sources.

For locating multiple light sources, "brainy" sensors can be used in two modes: (i) "localization" mode for localization and intensity estimation of light sources, when only a priori knowledge available is the number of light sources and (ii) "imaging" mode for estimation of intensity of the given number of light sources in the given locations, such as, for instance, in regular grid on a certain distance from the sensor. In the latter case, computational complexity and estimation errors, given the number of elementary sensors and their noise level, are very substantially lower. Figs. 3 and 4 illustrate these two operational modes.

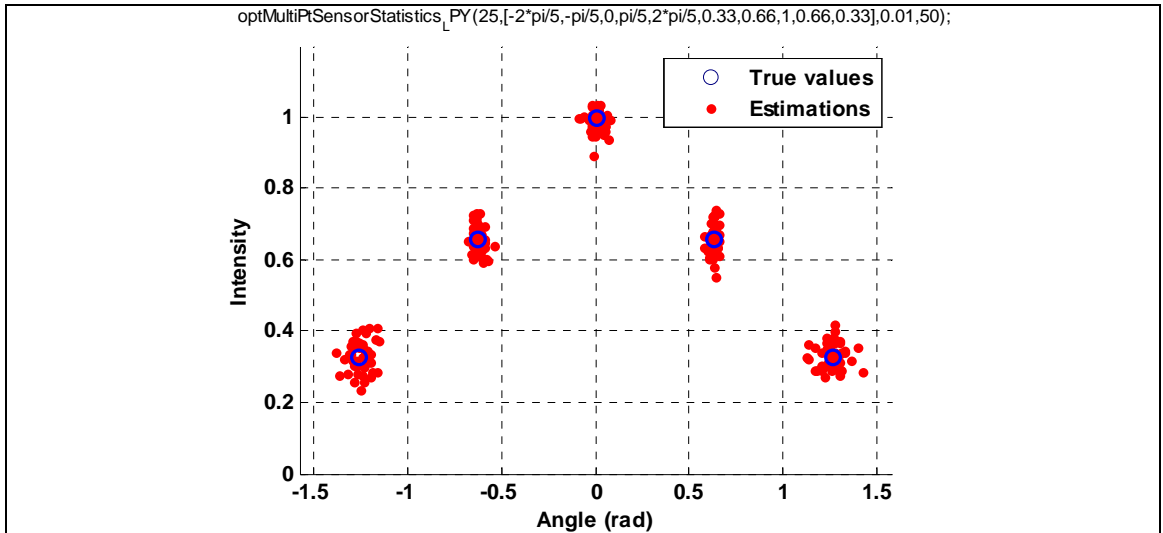

Fig. 2. Plots, in coordinates source intensity vs angular direction, of results of 50 statistical tests on localization and intensity estimation of 5 light sources by "brainy" sensor consisting of 25 elementary sensors with signal-to-noise ratio 100.

An important fundamental limitation of the "brainy" sensors is their threshold sensitivity and localization accuracy. If, for the given number of light sources, the number of elementary sensors and/or their signal-to-noise ratio approach certain threshold level, sensors' estimation errors grow very rapidly right up to complete loss of source detection capability.

## *Could nature do something like this? "Skin" (cutaneous) vision in nature*

It turns out that "skin" (cutaneous) vision is not rare among live creatures. Obviously, plants, such as sunflowers, that feature heliotropism must have a sort of "skin" vision to determine direction to sun and direct their flowers or leaves accordingly. There are also many creatures that have extraocular photoreception. For instance, cutaneous photoreception was found in reptilias ([9,10]) Annelid worms use quite ordinary camera-like image formation onto a quite ordinary retina. But they also have light detecting sensor arrays as well [11].

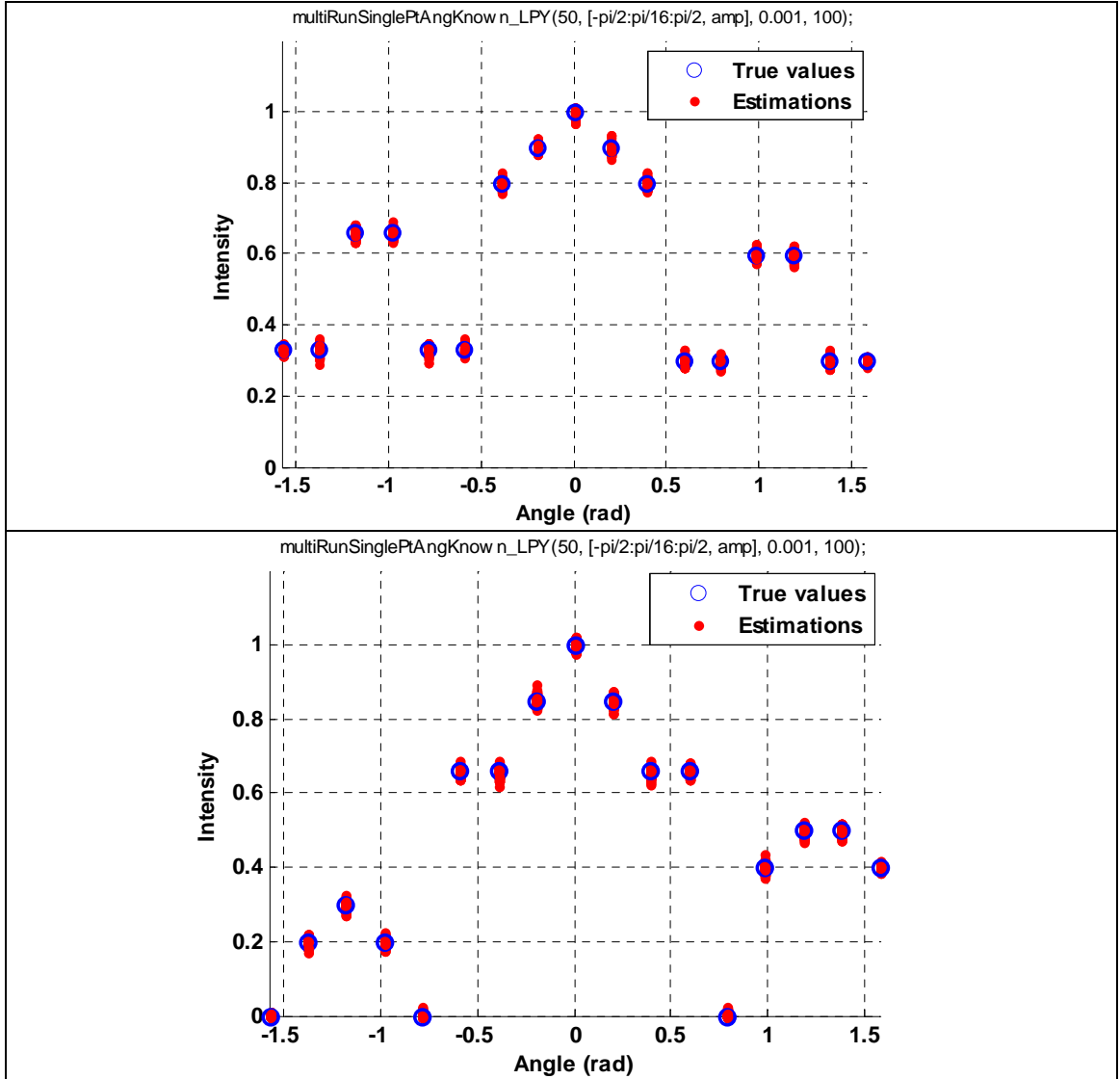

Fig. 3. Plots, in coordinates source intensity vs angular direction, of results of 50 statistical tests on intensity estimation of two patterns of 17 light sources with known angular directions by "brainy" sensor consisting of 25 elementary sensors with signal-to-noise ratio 100.

There is quite a number of reports on the phenomenon of "skin" vision in humans. Some of them have provoked skepticism ([12, 13]). However there are quite credible publications in favor of existence of this phenomenon as well. One of the most credible works is that by scientists in Russia M. M. Bongard and M.S. Smirnov ([14, 15]) who investigated famous Russian "medium" Rosa Kuleshova (one of the present authors, L.P.Y., witnessed seminar discussions of this work in the Institute of Information Problems of Russian Academy of Sciences). Here is a summary of their findings derived from carefully performed optometric and colorimetric experiments with Rosa Kuleshova.
- In general, the experiments revealed that Rosa Kuleshova did demonstrate the ability of "skin" vision with the fingers on her right hand

- Rosa Kuleshova demonstrated quite reliable ability of distinguishing illumination color with the fingers of her right hand. Her spectral sensitivity was in the range 0.42-0.68 mcm, the same as that of common human vision. From colorimetric experiments it was revealed that Rosa's "skin" color vision has, in her fingers of right hand, three types of receptors with spectral sensitivity similar to that of cones in human eye retina. Her left hand showed essentially no color "vision"
- The flicker fusion frequency of her "skin" vision was between 30 and 50 Hz, similar to that of common human vision
- The spatial resolving power of her finger vision was found to be about 0.6 mm, which corresponds to the density of skin light receptors about 10 per mm$^2$
- The sensitivity threshold of her "skin" finger vision to variations of intensity of illumination was about 10%. The sensitivity threshold decreased and resolving power increased when her fingers were moving across the target.
- Rosa Kuleshova was able to "see", with her fingers, large geometrical figures and other patterns on distances about 0.5- 1 cm but not more that 2 cm. For larger distances, Rosa's finger vision became very instable

Note that, according these data, total number of receptors in Rosa's fingers can be estimated as 1000 to 2000 assuming that the sensitive area of the fingers is about 1-2 cm$^2$. This is in a very god correspondence with above-outlined simulation results on "brainy" sensors. Threshold sensitivity property of "brainy" sensors may also explain low reliability of some experiments with Rosa Kuleshova. Decrease of the sensitivity threshold and increase of the acuity of Rosa's finger vision when her fingers were moving across the target can also be explained if one assumes that the neural circuitry, which processes elementary sensor signals, possess memory, and, therefore, movement of the sensors is equivalent to the increase of the number of sensors.

Yet another credible publication on finger vision is that by Zavala at al in [16]. More recently, an unequivocal, repeatable, high quality proof that a pattern of detections on the skin (abdomen, back, foot, and tongue have all been used) has been given by Paul Bach-y-Rita et al [17, 18] who use it to allow blind people to see, and so forth through what they calls "sensory substitution."

## *Imaging with the Lens in Contact with the Retina and intermediate cases*

An alternative approach to point light source location without an image involved placing the lens in direct contact with the detector array – something that precludes camera-like imaging. Could nature do something like that? Retinas in contact with a lens are found in a bioluminescent fireworm [19]

The pit for IR object detection in pit vipers is much too large to provide a useful image for these snakes. Yet they seem to have reasonably good prey and threat location ability. Recent work [20] has shown that computational imaging using a biologically plausible algorithm could produce a useful image form the detected image.

## *Conclusions*

> *There are more things in heaven and earth, Horatio,*
> *Than are dreamt of in your philosophy.*
>
> Hamlet by William Shakespeare

Once again, it seems that nature has anticipated human invention. There is enough circumstantial evidence presented here to suggest that computational imaging with data taken in ways other than camera-like systems may be going on in nature regularly. On the other hand, our "brainy" sensors show that quite good directional vision is possible even with sensors, whose angular selectivity is limited by the simple natural cosine law, and believe that they may cast a new light on evolution of vision, on mechanisms, advantages and limitations of such extra-ocular vision in nature and, perhaps, suggest new ways for planning experiments with animals and humans on further, more reliable and treatable study of the phenomena of extra-ocular vision.

An interesting question is why the Nature selected lens-based camera vision rather than "skin" vision as a main vision instrument in the animal kingdom? The answer might lie in very high computational complexity of the "computational" vision when high resolution within large field of view is required. What lens does in parallel and with a speed of light, computational vision must replace by computations in neural machinery, which is slower and require high energy (food) consumption. On the other hand, "computational" vision is very efficient and economic in detecting and localization of a single or some few light sources. We can only hypothesize that "skin", or cutaneous vision, apparently, appeared on much more early stages of evolution than camera-like vision, even earlier then it is commonly accepted ([2]), and then evolved, through sharpening of angular selectivity of elementary sensors, into compound eye vision, in which the number of light sources is equal to the number of elementary sensors and the computational complexity of neural machinery required is proportional to this number.